\documentclass[journal abbreviation]{copernicus2}

\usepackage{color}
\usepackage[T1]{fontenc}

\begin{document}

\title{Sub-ionospheric AKR: A possible mechanism for its transport down from the topside generation region to the F layer and ground
}

\author[1]{R. A. Treumann\thanks{Visiting the International Space Science Institute, Bern, Switzerland\\ \\ \emph{Correspondence to}: W. Baumjohann (Wolfgang.Baumjohann@oeaw.ac.at)}}
%\author[3]{R. Nakamura}
\author[2]{W. Baumjohann}
\author[3]{J. LaBelle}

\affil[1]{Department of Geophysics and Environmental Sciences, Munich University, Munich, Germany}
\affil[2]{Space Research Institute, Austrian Academy of Sciences, Graz, Austria}
%% The [] brackets identify the author to the corresponding affiliation, 1, 2, 3, etc. should be inserted.
\affil[3]{Department of Physics and Astronomy, Dartmouth College, Hanover NH 03755, USA}

\runningtitle{Sub-ionospheric AKR}

\runningauthor{R. A. Treumann, W. Baumjohann and J. LaBelle}

%\correspondence{R. A.Treumann\\ (art@geophysik.uni-muenchen.de)}

\received{ }
%\pubdiscuss{ } %% only important for two-stage journals
\revised{ }
\accepted{ }
\published{ }

%% These dates will be inserted by the Publication Production Office during the typesetting process.

\firstpage{1}

\maketitle

\subsection*{Abstract}
%\begin{abstract}
Recent theoretical work shows that "electron exhausts", elongated regions of depleted electron density shown by PIC codes to be generated in collisionless reconnection, can in principle become sources of X-mode Auroral Kilometric Radiation generated by the electron cyclotron maser instability (ECMI). In this paper it is suggested that under certain conditions such X-mode AKR can possibly be transported down along the auroral magnetic flux tubes to low altitudes, in extreme cases even penetrating below the ionospheric F- and E-regions to reach ground level. The characteristic features of low-altitude AKR produced and transported by this mechanism would be bandwidths of order $\sim$100 kHz and occurrence in quasi-periodic bursts lasting $< 1$ s when passing at full Alfv\'en speed, $>1$ s when retarded. Most published observations of low-altitude AKR have insufficient time resolution to detect these features, although a few are suggestive of them. The few observations with marginally sufficient time resolution are so far neither consistent nor inconsistent with these predictions, suggesting that occasions when electron exhausts transport AKR all the way to low altitudes may be rare. However, vigorous experimental effort to test these theoretical predictions is warranted because of its physical and astrophysical significance. They would also provide an additional method to remotely detect and investigate reconnection physics.

%\end{abstract}

\vspace{0.5cm}
\section{Introduction}
Auroral kilometric radiation (AKR) was first observed by \citet{benedictov1966}, \citet{dunckel1970} and \citet{gurnett1974} with the last of these clearly identifying its significance, power level, and origin in the upper auroral ionosphere \citep[see also][for a systematic review of observations]{gurnett1983}.  It is the most powerful natural radio radiation from Earth, comprising at times up to 1\% of the energy of the aurora. 

AKR is generated by the weakly-relativistic electron-cyclotron-maser instability (ECMI) which excites all four modes, the lower-branch X-mode, low-harmonic upper-branch X-modes, O-modes and Z-modes though at rather different growth rates. The ECMI was first recognised in the possible negativity of the absorption coefficient in dilute plasma conditions  \citep{twiss1958,twiss1958a,hirshfield1964}. \citet{melrose1976} applied it to explain observations on Jupiter and in the aurora. Unreasonably large temperature anisotropies were needed, however, which led him to favour gyro/synchrotron emission instead. Success of the ECMI started with \citet{wu1979} who realised the importance of relativistic effects in the resonance condition. Models of AKR generation via ECMI are well-developed \citep [see][for systematic reviews of the ECMI]{melrose1986,melrose1991,melrose2017,treumann2006}. The predominant emission is X-mode radiation emitted outward and observable only from space.

However, over the years observations have suggested that a component of AKR penetrates to low altitudes. For example, recent ground-based observations in Antarctica show occurrence of AKR-like signals simultaneous with similar AKR detected in space with the Geotail spacecraft \citep{labelle2011a,labelle2015}. These observations confirmed earlier claims that AKR might be observable at low-altitude spacecraft, rockets and possibly even from ground-based instrumentation \citep {oya1985,morioka1988,beghin1989,yoon1995,shutte1997,labelle1999,parrot2012,simoes2012}. Such observations are difficult to explain by conventional mechanisms because the highly ionised dense upper ionosphere poses a seemingly unsurpassable barrier for the transport of X-, O-, or Z-mode radiation at AKR frequencies of a few 100 kHz down to the atmosphere. \citet{oya1985} proposed mode transformation from O-mode AKR to whistler mode which can under certain conditions penetrate to low-altitude or even ground level in ducted or non-ducted propagation modes. Transformation of the parallel propagating O-mode to the whistler mode at constant frequency implies that the latter becomes an electron Alfv\'en wave with wavenumber $k_\|\sim \lambda_e^{-1}$ matching the inverse electron skin depth. In order to cross the E-layer downwards this requires another transformation of the whistler back to the O-mode. This model, or some variation of it, holds promise to explain many of the observations. Other, more exotic mechanisms have also been proposed, such as ballistic wave transformation \citep{krasovskiy1983}. 

None of the mechanisms so far involves \emph{direct transportation} of AKR to low altitudes. Such direct down-propagation requires low densities in the radiation-carrying magnetic flux tube, corresponding for example to some kind of perforation of the ionosphere. Though not impossible, these are difficult to imagine on their own and require a more sophisticated mechanism as one can hardly, if at all, expect that entire flux-tubes along the auroral geomagnetic field over their complete length could be nearly depleted from plasma. 

Another approach to direct down-propagation involves electron holes transporting X-mode radiation downward along the magnetic field \citep{treumann2012}. Electron holes, being Debye-scale objects, cannot support excitation of the ECMI at kilometre-long wavelengths, but this difficulty can in principle be overcome in a statistical approach since observations in the upper auroral ionosphere have proved the existence of myriads of Debye-length scale electron holes which could, if distributed over one AKR wavelength, piecewise amplify AKR in a statistical sense, a model which has not yet been investigated in any depth. However, although amplification of AKR might be possible, trapping of radiation and transport to low altitudes in electron holes appears impossible, even statistically.

A third possibility for direct down-penetration is inspired by 
recent work  \citep{treumann2017} showing that ECMI can operate to produce X-mode AKR in so-called "electron exhausts", elongated regions of vastly depleted electron density generated by collisionless reconnection as well established in numerical particle-in-cell simulations \citep[cf., e.g.,][for an overview]{karimabadi2013}. The reconnection-generated electric field transforms the dilute trapped electron distribution into a strongly deformed shape. In a strong guide field the whole process is strictly two-dimensional (2-D) with the exhaust becoming asymmetric, and the deformation of the distribution function is in the momentum space perpendicular to the guide field. Since the exhausts are spatially extended in all three dimensions, these conditions are strongly in favour the ECMI. Fundamental radiation is just beneath the electron cyclotron frequency $\omega\lesssim\omega_c=eB_g/m$ which in the strong auroral geomagnetic field is the observed AKR frequency range. In a region where the plasma density is not too high, as for instance in the auroral ionospheric trough, low cyclotron harmonics, if amplified by the ECMI could exit from the exhaust to free space. In general a mechanism of this kind is of high interest in auroral, planetary, solar, and stellar physics as even in astrophysics because of its potential applicability to the radio-radiation emitted from the sun, other magnetized stars, strongly magnetized objects (white dwarfs,  pulsars) or also extended turbulent plasmas.

Below, we investigate the alternative possibility whether AKR generated and trapped within electron exhausts associated with reconnection in the nightside magnetotail can be transported downward to low-altitudes and even occasionally penetrate the lower ionosphere to release the radiation into the free non-conducting atmosphere where it can be observed at ground-level.

\section{The model}

The reconnection-generated ECMI model parallels the electron hole-generated ECMI \citep{treumann2011,treumann2012}, in that reconnection exhausts function like large electron holes. They are to a large degree ($\sim 90\%$ ore more) depleted from electrons, and they contain an anisotropic trapped dilute weakly relativistic electron distribution which serves as the radiation energy source. Because the source is localised, the ECMI mechanism can be treated in the local frame of the exhaust which simplifies the calculations. Furthermore, the radiation source (exhaust) can be treated as circular for purposes of calculating the resonance condition, although the circular simplification is not crucial since inclusion of an eccentricity just complicates the calculation, affects the amplified bandwidth, but does not change the essential physics. 

The two main properties in favour of the ECMI are the presence of the strong current-aligned guide field and the spatial extent of the exhaust which is of the order of hundreds of electron inertial lengths $\lambda_e=c/\omega_{ea}$ in the two directions perpendicular to the current and guide field, where $\omega_{ea}$ is the ambient plasma frequency outside the exhaust. Along the guide field, the exhaust extends several ion inertial lengths $\lambda_i=\sqrt{m_i/m_e}\lambda_e$, which amounts to $\sim 100\lambda_e$. The ratio of electron inertial to Debye length is $\lambda_e/\lambda_D=c/v_e \gg 1$ for any reasonable auroral electron thermal speed $v_e$. Thus reconnection exhausts, in contrast to electron holes, are huge objects which can contain at least one unstable AKR ECMI wavelength. Conditions inside the exhausts are ideal for wave growth via ECMI, fed in energy by the perpendicular momentum space anisotropy of the trapped weakly relativistic reconnection generated electron distribution. The growth rate is proportional to the internal plasma to cyclotron frequency ratio $\omega_e/\omega_c$, with $\omega_c=eB_g/m_e$ mainly determined by the strong guide field. The low value of this ratio suggests that the amplification might be weak. However, any trapped ECMI wavelength which is excited at frequency $\omega\lesssim\omega_c$  has many amplification cycles available in order to grow to a substantial amplitude as long as the trapping conditions hold inside the exhaust and it cannot escape. The same applies to any trapped lower cyclotron harmonic, a situation that strongly favours excitation and amplification of the ECMI.

The expression for the maximum growth rate of the wave with frequency $\nu=\omega/\omega_c$ following from the ECMI theory under the conditions for reconnection is \citep{treumann2017}
\begin{equation}
\frac{\gamma(\omega)}{\omega_c}\approx \frac{\alpha\pi^3}{4\nu} \frac{\omega_{ea}^2}{\omega_c^2}\bigg(\frac{V_A}{v_e}\bigg)\bigg(\frac{c}{v_e}\bigg)\exp\bigg(-\frac{1}{2}\frac{v^2_e}{V_A^2}\bigg)
\end{equation}
with $\alpha=N_h/N$ the ratio of exhaust-trapped to ambient densities. This ratio is of the order of $\alpha\lesssim \mathrm{O}\,(0.1)$.
This expression holds for the fundamental and its harmonics $\nu\approx 1,2,3\dots$. Thus the growth rate is determined essentially by the ratio $\alpha\:\omega_{ea}^2/\omega_c^2$ and the exponential factor. It decays with harmonic number $\nu$. Note that $v_e$ is the thermal speed of the hot and reconnection-accelerated trapped electrons which can be substantial, of order $v_e\gtrsim 10^4$ km/s. With $V_A\approx 2\times 10^3$ km/s in the exhaust, the ratios $V_A/v_e\lesssim 0.2,\ c/v_e\lesssim 30$ imply a growth rate
\begin{equation}\label{eq-gamma}
\gamma(\omega)\lesssim 5\times 10^{-3} \alpha\frac{\omega_c}{\nu}\bigg(\frac{\omega_{ea}}{\omega_c}\bigg)^2
\end{equation}
sufficiently small to compensate for the ambient plasma to cyclotron frequency ratio, in particular for large depletions $\alpha\ll 1$. Nonetheless, proportionality to the ambient field cyclotron frequency makes this a substantial growth rate. It implies a growth time $\tau=\gamma^{-1}$ of some $10^3<\tau \omega_c<10^5$, corresponding to a few thousand cyclotron times, probably long enough for reaching quasilinear saturation in the trapped radiation intensity, which grows exponentially according to
\begin{equation}
I^\mathrm{X}_{\omega_k,k}(t)\propto I_{\omega,k}^\mathrm{th}\exp[2\gamma(\omega)t],
\end{equation}
until quasilinear saturation at time $t=\tau_{ql}$ when the momentum space gradient of the electron distribution becomes depleted. $I^\mathrm{th}_{\omega,k}$ is the thermal level of the X mode from which it starts growing. Relevant thermal levels for the instability to start from have recently been calculated \citep{yoon2017} numerically. 
 
The ECMI excites propagating X and O modes, as well as Z modes \citep[cf., e.g.,][]{yoon1998,yi2013}. The focus herein is on the X mode. It propagates perpendicular to the magnetic field, which in the case of reconnection exhausts is the guide field $\vec{B}_g$,  the ambient geomagnetic field. The fundamental wave is excited below $\omega_c$ and thus on the lower X mode branch below the free space cut-off, implying that it is trapped within the exhaust. Harmonic waves could escape provided the harmonic number of the excited wave $\nu$ exceeds the ambient X mode cut-off ratio
\begin{equation}\label{eq-4}
\nu> 1+\omega_{ea}^2/\omega_c^2
\end{equation}
It is thus clear that the radiation excited inside the exhaust cannot escape unless excited at sufficiently high harmonic number. If the ambient plasma to cyclotron frequency ratio is small, the second harmonic should be enabled for immediate escape from the exhaust but will be weak due to experiencing just one or even only a fraction of growth time. 

\section{Discussion}

ECMI-excited waves at the fundamental or sufficiently low harmonics, depending on gyro- to plasma frequency ratio, become trapped through reflection from the boundaries of the reconnection exhaust structure, allowing them to become amplified until reaching the quasilinear equilibrium \citep{yoon1995} between the trapped electron distribution function and the growth of the wave, as well as perhaps being weakly Landau damped. Typical quasilinear saturation times are of the order of $\tau_{ql}\gtrsim 100\tau$ thus amounting to $10^5<\tau_{ql}\omega_c<10^8$. Therefore, for a nominal AKR frequency of $\omega/2\pi\sim 300$ kHz, quasilinear saturation would set on after $10^{-3}<\tau_{ql}< 1s$. Since quasilinear saturation implies depletion of the perpendicular momentum gradient in the electron distribution which excites the ECMI, this time determines the height range over which the waves grow. 
Assuming that the exhaust moves along the magnetic field at a speed close to the Alfv\'en velocity $V_A$, as argued below, the height range is $\Delta h \approx \tau_{ql}V_A$, which amounts to $1< h<10^3$ km for $V_A\sim 10^3$ km/s. In the dipole magnetic field this fixes the frequency range to $\Delta\omega\lesssim 0.47\omega_c$ which, for $\omega_c/2\pi\approx 300$ kHz, corresponds to a bandwidth $\lesssim 140$ kHz assuming fundamental radiation, partially bridging the gap to the next trapped harmonic.

Trapping of radiation also allows for the evolution of nonlinear effects other than quasilinear saturation among the trapped ECMI modes. The trapped waves will also interfere with each other, implying challenges for observational mode-identification. Constructive and destructive interference produce spatial fringes of the emission which are converted into frequency modulation through the relative motion of the spacecraft and exhaust structure. For example, consider a single wavelength undergoing a single reflection through an angle $\theta$. The spatial separations of the resulting constructive interference maxima are $d=\lambda/|\sin\theta|=2\pi c/\omega|\sin\theta|$. The pattern moves at Alfv\'en speed, thus yielding an apparent frequency modulation in the spacecraft frame of $f\sim V_A/d= 2\pi V_A\omega|\sin\theta|/c\lesssim$ few kHz. Multiple reflections would produce more complicated patterns, in general non-sinusoidal and not of equal amplitude. The resulting quasiharmonic modulation spectrum $nf,\ n=1,2,3\dots$ may resemble a fingerprint of spectral extension $\lesssim\Delta\omega/2\pi$.

Transport of the ECMI-generated AKR trapped in the reconnection exhaust occurs via the spatial translation of the exhaust structure. \citet{treumann2017} show that in the auroral zone, where the current is carried by kinetic Alfv\'en waves, respectively inertial Alfv\'en waves (IAW) whose perpendicular scale is the electron inertial length $\lambda_e$ based on the ambient plasma density, it is the magnetic field of the IAW which undergoes reconnection. This wave \citep[cf., e.g.,][for the the linear and nonlinear theory of kinetic Alfv\'en waves]{yoon2017a}, by current understanding, flows mainly upward in the upward current region and downward in the downward current region. Hence the exhaust structure generated in the reconnection process propagates with the IAW, either upward or downward depending on its direction of propagation.

Trapping in down-going exhausts is particularly effective for at least two reasons. First and most important, the ambient density increases with decreasing altitude implying that the cut-off ratio Eq. (\ref{eq-4}) increases.  This can be written
\begin{equation}
\nu> 1+(m_e/m_i)c^2/V_A^2
\end{equation}
with $V_A$ strongly decreasing when approaching the F-layer. The frequency of the trapped AKR remains constant below the relatively high altitude at which the quasilinear state is reached, turning off the excitation of higher frequencies at lower altitudes. Second, because the ambient magnetic field increases with decreasing altitude, for the downward moving case the frequency of the saturated waves becomes much lower than the local cyclotron frequency. Hence, downward-moving exhausts trap the internally-amplified frequency interval $\Delta\omega$ of the fundamental, second or even third ECMI-AKR harmonic radiation leading to effective confinement over the entire distance over which the exhaust structure survives.

A significant consequence of the role of IAW in reconnection and transport of the exhaust structures is that, since the IAW magnetic field is quasi-periodic with wavelength $\lambda_\|\approx 2\pi/k_\|$ along the geomagnetic guide field, one expects a chain of independent reconnection exhausts to be produced, with all exhausts moving together with the wave up or down the ambient field. Its field-aligned velocity is approximately of the order of the inertial Alfv\'en speed, roughly a factor two less than the nominal Alfv\'en speed, thus being of the order of several 100 km/s. PIC simulations of reconnection \citep[cf., e.g.,][]{jaroschek2004} suggest that, in general, exhausts expand and shift in the direction perpendicular to the ambient guide field at individual speeds. In an extended (say plane) current sheet they interact with each other; for example, some small exhausts can be destroyed or "eaten up" by larger ones. It is not precisely known what determines their individual velocities. There may also be differences in their parallel velocities caused by neighbouring exhausts along the ambient field. Therefore, a picture which refers to a common constant speed $\sim V_A$ must be considered as approximate. These differences between different exhaust structures, as well as their evolution and interactions, will of course be reflected in the complex frequency-time spectra of the trapped ECMI radiation and also in any escaping radiation.
\begin{figure}[t!]
\includegraphics[width=0.5\textwidth,clip=]{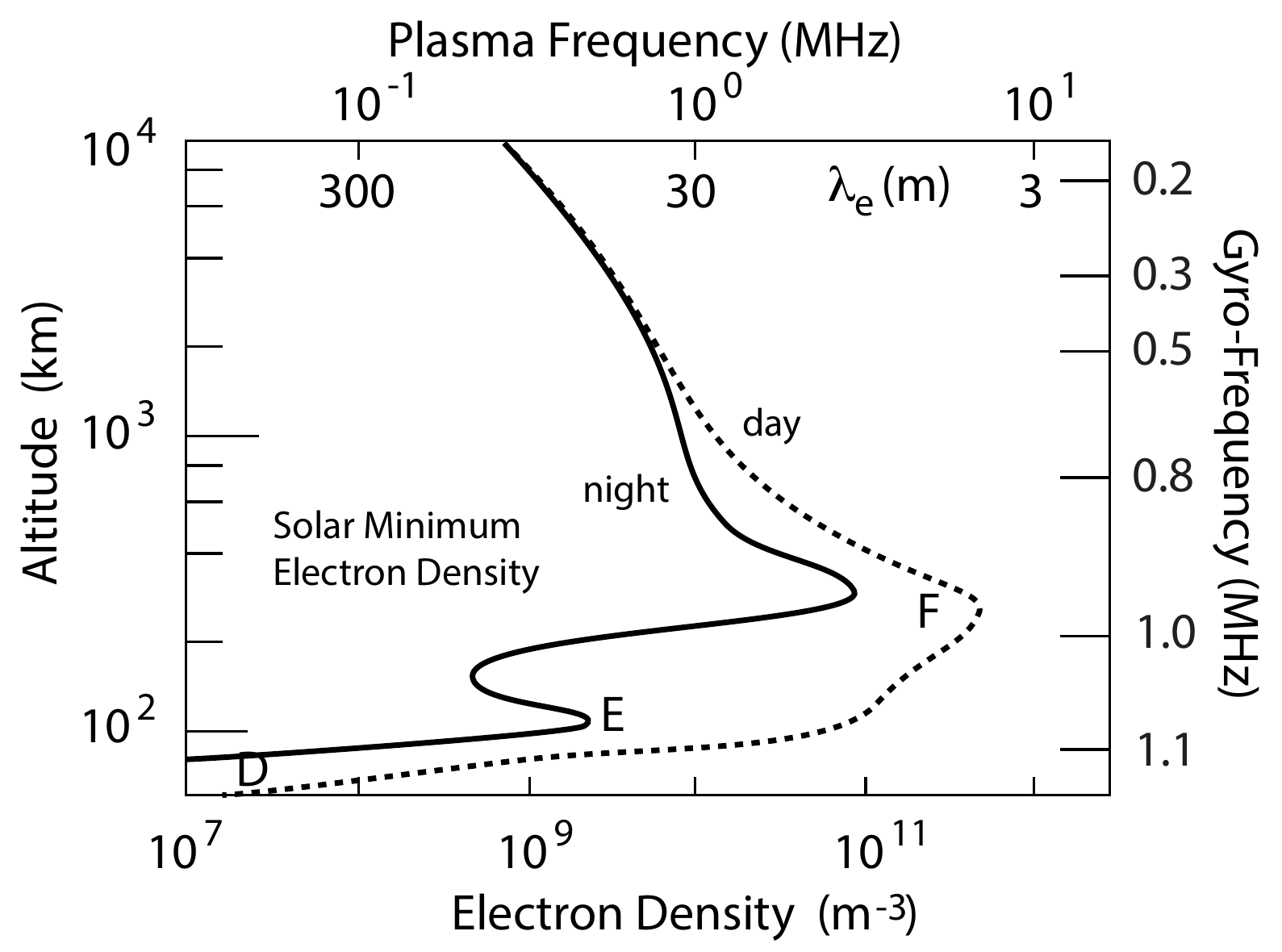} 
\caption{A summary of ionospheric electron density profiles during solar minimum below $10^4$ km altitude \citep[adapted from][for solar minimum conditions]{kelley2009}. Only nighttime densities are of relevance for AKR-ECMI. The ionospheric F-, E-, and D-layers are indicated. The right ordinate shows the variation of the nominal geomagnetic electron-cyclotron-frequency with altitude as given on the left ordinate. The upper abscissas show the scales of the ambient electron plasma frequency $\omega_{ea}(h)/2\pi$ and the related ambient electron inertial length $\lambda_e$. }
\label{fig-akr1}
\end{figure}

PIC simulations have so far only followed the evolution of reconnection exhausts for short times and distances, insufficient to observe their decay. Hence their lifetime is not known. Presuming they last long enough, the displacement time of an exhaust structure from the AKR source region at around $<3$ Earth radii altitude to the ionosphere would be of the order of $10-20$\,s, corresponding to a  large number of cyclotron periods of the trapped quasi-linearly saturated wave. According to Eq. (\ref{eq-gamma}) and assuming an AKR frequency $\omega/2\pi\sim 300$ kHz this number is $>10^8-10^9$.

The question now arises under what condition the waves can become de-trapped from the exhaust structure and escape into the ambient medium. Two mechanisms of escape are considered here: tunnelling or leaking out of the exhaust structures as they move along the field, and escape of radiation when the structures dissolve should they survive long enough to reach the bottom of the ionosphere. The former process depends to some extent on whether the trapped waves remain perpendicular to the ambient magnetic field or refract to the parallel-propagating condition. In either case, the trapped X mode can propagate inside the exhaust as long as its wavelength fits into the exhaust and its frequency stays above the lower cut-off of the lower branch of the X mode. For perpendicular propagation de-trapping (escape) is possible only in regions where $\omega_{ea}<\omega_c$. For an AKR frequency of, say, 300 kHz, this condition requires that the nighttime auroral density be less than $N\lesssim 7.2\times10^8$ m$^{-3}$, which is far lower than the ionospheric E- and F-region densities even under nighttime solar minimum conditions \citep[see Fig. \ref{fig-akr1}, and][]{kelley2009}. Thus, as mentioned above, such waves will be trapped over the entire distance over which the exhaust structure survives.

In case the scattering of the trapped AKR wave within the boundaries of the exhaust structure has gradually turned the direction of propagation from perpendicular into parallel, the wavelength-fitting criterion for trapping is more easily met because the exhaust dimension parallel to the ambient field is of the order of the ion inertial length, which is  roughly 10 to 100 times longer than the perpendicular size. The trapped parallel AKR can propagate on the parallel lower branch R mode. On this branch, when escaping, it can convert to the whistler mode and propagates. Whistler modes converted from the ECMI-generated waves refracted into parallel conditions may therefore explain the occasional observation of whistlers under AKR conditions at upper auroral ionospheric altitudes.

A key characteristic of the reconnection exhaust structures, which has significant consequences for radiation trapped within them, is that they shrink as they propagate downwards. This occurs due to the increasing external pressure on the exhaust. The initial size of the exhaust structure is critical for determining the trapping, or lack thereof, of the ECMI-generated waves as the structure moves to low altitudes. If the initial structure is too small, once the ambient $\lambda_e$ matches the trapped ECMI-AKR wavelength and small exhausts have shrunk to this size, the AKR X mode wave cannot be contained anymore and should leak out. However, at  altitudes where the ambient density is high such that  $\omega_{ea}>\omega_c$, the X-mode on the lower branch cannot propagate and will become evanescent and absorbed by the surrounding plasma. In such small dying exhausts this is the fate of the X mode if it is not turned into an R mode by multiple reflections, as claimed above.

An idea of the trapping capacity of the exhausts can be obtained when considering the evolution of the ambient $\lambda_e$ during descent into the F layer. This  electron inertial length is inversely proportional to the root of the ambient density $N$. Hence, a density increase by  a factor of 100 during downward motion from the AKR  source to the F-layer implies a decrease of the inertial length by a factor 10, viz. $\lambda_{eF} \to 0.1\lambda_e$ which causes the exhaust to shrink. As a consequence the number of trapped wavelengths is reduced by roughly a factor 10, while the trapped bounce frequency increases.

Figure \ref{fig-akr1} gives an idea of the variation of the plasma frequency and the related ambient electron inertial lengths with decreasing altitude. From the scale of $\lambda_e$ and the nighttime profile during solar minimum, when the ionospheric densities are lowest and thus favourable for our process, one realises that electron exhaust structures must in the magnetosphere have scales of around several $\sim10^2\lambda_e$ perpendicular to the ambient geomagnetic field in order to survive the shrinkage they experience when entering into the nighttime F-layer to maintain at least one single AKR wavelength. This is not impossible, however, for the largest electron exhausts which form in reconnection at higher altitude. Such exhausts may extend  several $10^2\,\lambda_e$ to $>10^3\lambda_e$ and may therefore in some rarer cases survive even transport across the nighttime F-layer, which has peak densities of $N\sim 10^{10}-10^{11}$ m$^{-3}$ corresponding to ambient plasma frequencies $\omega_{ea}/2\pi\approx 3$ MHz and cyclotron frequencies $\omega_c\sim 0.9$ MHz.  Assuming that the Alfv\'en velocity does not change substantially (but see the discussion below), transport across the F-layer takes $(0.1-0.2)$ s and is followed by encountering the E-layer at $120-100$ km altitude. 

The crossing time of the E-layer for large exhausts will be short enough to occur with little damping, and upon encountering the underlying  D-region any exhaust will immediately dissolve, as the ambient plasma density drops to negligible values, and the trapped radiation is released. Practically no ambient plasma density is present at such low altitudes. Radiation escaping below the E-layer is exposed to the practically neutral atmosphere which still contains a dilute electron population. It propagates as ordinary free-space electromagnetic radiation with either X mode or R polarisation depending on whether it remained in the X-mode until the end, or whether it was scattered/refracted into parallel propagating R-mode whistlers. In this case, the resulting radiation could even be detected at ground-level.
\begin{figure}[t!]
\centerline{\includegraphics[width=0.5\textwidth,clip=]{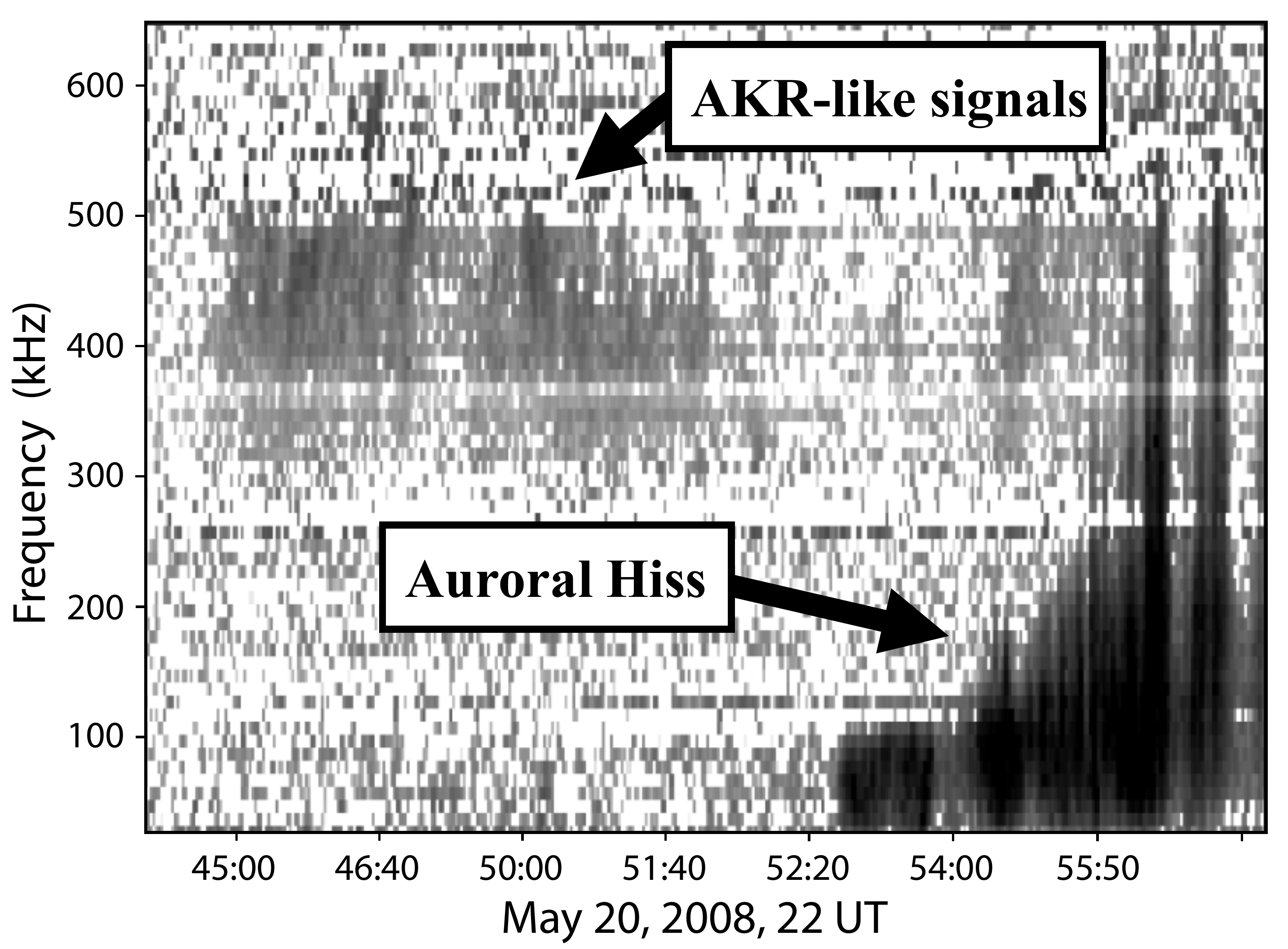} }
\caption{Roughly 11 minutes of several hour long observations of natural radio spectra in the spectral range of AKR at South Pole Station from May 20, 2008 \citep[an example from the large data base studied by][]{labelle2015}. Here AKR-like signals are seen for about 5 minutes between 300-500 MHz, exhibiting upward and downward frequency drifts. Time resolution at ground was $\lesssim$ 2 s which did not allow resolving sub-second variations thus inhibiting identification of single exhausts moving at  Alfv\'en speed. This downward South Pole AKR activity coincides in time with the upward AKR recorded on the Geotail spacecraft \citep{labelle2011}. } \label{down-fig-1}
\end{figure}

The least dense and smallest exhausts will decay early up in the ionosphere. Actually, already in the F- and in particular the E-layer the IAW current deviates from being field-aligned and in the E layer in the presence of onset of perpendicular collisional conductivities starts closing in the perpendicular direction. Therefore exhausts already start decaying and dissolving at these altitudes. Decay of large  exhausts below 1000 km altitude, and release of the trapped AKR mainly in the whistler mode may explain some of the low altitude satellite and rocket observations of AKR-like radiation.

Turning to the data, reported observations 
of AKR-like events at low altitudes with low Earth orbit satellite, rocket, or ground-based instruments are sparse, and most of them are either of short duration or have time resolutions of order one second or longer. Because the travel time of IAW from source to ionosphere is greater than or not much less than the cadence of these measurements, this resolution is generally insufficient to resolve the expected bursts or quasi-periodicity of AKR transported in electron exhausts. Nevertheless, some of these observations are suggestive of the kind of bursty structure expected from this mechanism, for example sporadic short events reported by \citet{labelle2011}. 

Figure \ref{down-fig-1} from the data base studied by \citet{labelle2015} shows an example of  AKR-like signals observed on the ground at South Pole. In this example from May 20, 2008 the emissions exhibit a drift in frequency either upward or downward, and the stronger ones have bandwidth $>100$\,kHz. Drifts in these features might reflect the different frequency ranges trapped in the chain of exhausts with no preference for whether higher or lower frequencies arrive first. The recordings on this day also contain shorter less intense bursts. Unfortunately the $\lesssim$2-s cadence of these measurements makes it impossible to verify that time durations of the features match to expected IAW scale sizes or travel times. A handful of measurements in the literature have finer time resolutions up to milliseconds \citep[see, e.g.,][who reported DEMETER spacecraft observations related to AKR]{parrot2012}; among these, one example comes from an auroral sounding rocket \citep{labelle1999}, another from South Pole Station
\citep[][their Figure 1]{labelle2015}. In both of these examples the AKR fine structure consists of many overlapping features with few-kHz bandwidths coming and going but continuously existing over an event duration of tens of seconds or more, closely resembling the fine structure of X-mode AKR observed by distant satellites sensing up-going AKR. These particular observations do not match characteristics expected of AKR transported in reconnection exhausts, but they are so few in number, it is not possible to exclude that a significant fraction of AKR-like events might have such structure. More high-resolution measurements are needed to answer that question.

\begin{figure}[t!]
\includegraphics[width=0.475\textwidth,clip=]{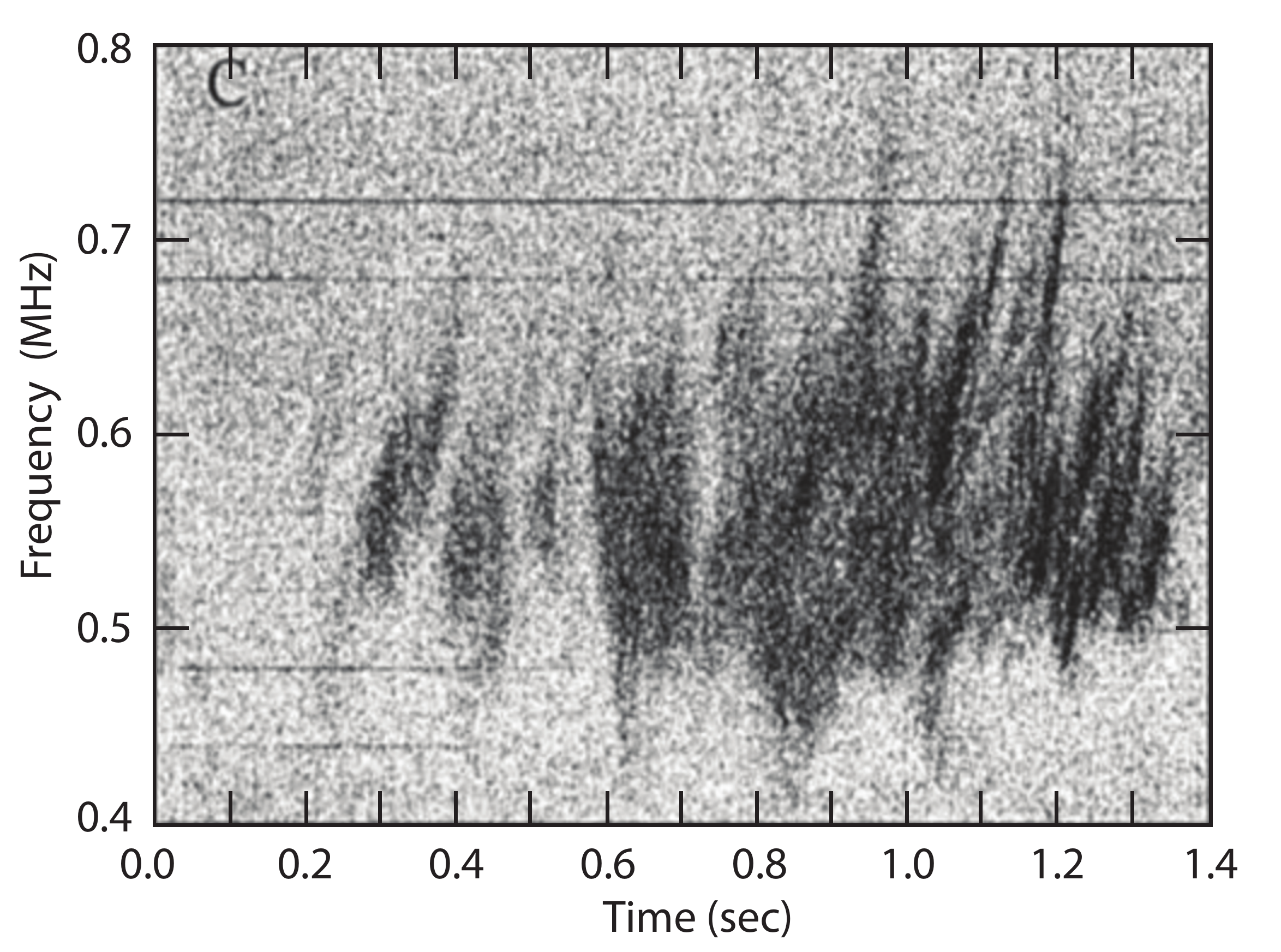} 
\caption{Sample spectrogram measured with the PHAZE II sounding rocket between 316 and 331 km altitude \citep[adapted from][Fig. 1c, see description in text]{labelle1999}. The grey scale is 20 dB. The visible irregular drifting structures are in the whistler band. They exhibit upward and downward drifting short emissions of comparably narrow local bandwidth. }
\label{fig-akr2}
\end{figure}

{It is of particular interest to discuss the observations of \citet{labelle1999} which were performed in situ the F-layer at 331 km altitude under over-dense plasma conditions in February 1997 near the Solar Sunspot Minimum. The ambient plasma frequency was about $\sim 1.7$ MHz, at cyclotron frequency $\sim 1.4$ MHz, a relatively small difference.  Figure \ref{fig-akr2} shows an example of their observations of irregular electromagnetic emissions which were observed sporadically between $0.4\lesssim \omega/2\pi\lesssim 0.8$ MHz in the AKR frequency range. These were interpreted as propagating on the whistler mode assuming that wave propagation was parallel to the ambient magnetic field. The patches exhibit both upward and downward drifts in frequency. It is not known how such whistlers would be generated locally. If interpreting them in the frame of the exhaust picture they could have resulted from exhausts which cannot pass the F-layer, decay and release the trapped AKR which has been transported down. This becomes possible when the wave vector of the trapped radiation has, by multiple scatterings, turned into nearly parallel direction,  and the radiation has changed from lower-branch perpendicular X to parallel R mode. After destruction of the exhausts the radiation would then be released on the ambient whistler branch at nearly parallel/antiparallel propagation with no preference of direction up or down along the magnetic field, because stochastic scattering inside an exhaust has no directional preference.} 

{The direction of the R-whistler mode drift may in addition be masked by the dynamics of the decay of the exhaust. If the exhaust simply shrinks in dimension then the longest wavelengths - lowest frequencies - should be released first. The spectral drift is then upward, and the bandwidth maps the bandwidth of the trapped radiation. This is overlaid over the internal propagation of the trapped R mode which in other cases partially or totally compensates for the frequency drift or even turns it into downward direction. From this point of view it is also interesting that the bandwidth of $\sim 100-200$ kHz of the observations is not in disagreement with the assumption of trapped AKR respectively R modes and their above estimated bandwidths. The total duration of a single drifting emission stroke in this case is a measure of the final local velocity of the exhaust. From Figure \ref{fig-akr2}  \citep[Fig. 1c in][]{labelle1999} one may estimate that the passage across one upward drifting structure takes about $\sim 0.1$ s. Assuming that the size of the exhaust has at this altitude shrunk to $\sim 1$ km for releasing the trapped radiation, the final speed of the exhaust has become small, just $\sim 10$ km/s, much less than the initial Alfv\'en speed, not contradicting the assumption that most exhausts terminate above or in the F-layer when moving down, decay, and release the trapped radiation. The transition time of a few kilometre long exhaust across a spacecraft is definitely too short to be detected. But when decaying, the released radiation in the whistler band propagates along the field and can become observed. Groups of emissions in such a case might indicate presence of a chain of decaying exhausts near spacecraft altitude. Rarely, a few large exhausts may possibly make it down to below the F-layer.}

Another question that arises from the possible explanation of ground-level AKR observations as resulting from extraordinarily large reconnection exhausts penetrating to below the F- and E- regions is why these have not been detected by incoherent scatter radars. Such radars easily detect, for example, rising bubble structures associated with Rayleigh-Taylor instability operating in the ionosphere. Such bubble structures often have dimensions 10 km or more, however, which is about an order of magnitude larger than the expected size of the reconnection exhausts which reach the F-region. Indeed from Fig. \ref{fig-akr1} at F-layer peak one has $\lambda_e\gtrsim 10$ m and thus an extension of an exhaust roughly of the order of $\gtrsim 1$ km, the minimum size for trapping the AKR at the fundamental of the lower X mode branch. {The reason for missing the exhausts might possibly be found in the comparably low thermal fluctuation level of Langmuir waves in the exhaust. Its ratio to the ambient thermal level is
\begin{equation}
\frac{\langle E^2\rangle_h}{\langle E^2\rangle_0}= \Bigg(\frac{N_h}{N_0}\Bigg)^\frac{3}{2}\sqrt{\frac{T_e}{T_{eh}}} \sim 10^{-2}
\end{equation}
Due to the low exhaust-electron density $N_h\sim 0.1N_0$, Langmuir fluctuations inside the exhaust are well below their ambient fluctuation level. The temperature increase of the dilute trapped electron component does not compensate for the density depletion. At F-layer altitude, the comparably small exhaust structures containing the estimated low electric fluctuation levels probably escape recognition by the incoherent scatter radar. On the other hand, each exhaust can be understood as a few kilometre scale low frequency density perturbation which, when many exhausts occasionally accumulate in the F-layer, could cause a signal in the backscattered ion line. It would be interesting to search for typical signals in relation to ground-based, rocket or low-orbit spacecraft observations of emissions related to AKR.}

{Substantial electron acceleration in the reconnection takes place when the energetic electrons which are released from one exhaust encounter the many neighbouring exhausts along and perpendicular to the ambient geomagnetic field. This acceleration is well documented in the many PIC simulations of reconnection \citep[cf., e.g.,][and others]{jaroschek2004}. In the auroral ionosphere the electrons released from each exhaust become accelerated and add up to contribute a high-energy tail to the field-aligned momentum  distribution of the energetic auroral electron population which causes the observed aurora and its related effects. These electrons cannot be distinguished in the auroral electron fluxes.}

One can also ask whether the sides or front edge of the exhaust structures would have sufficiently steep density gradients to drive instabilities producing structure that would light up coherent scatter radars such as the SuperDARN radars which probe large volumes of the high-latitude ionosphere nearly continuously. (Such instabilities would need to be strong to operate on the short time that it takes the exhaust structures to penetrate the ionosphere.) Of course no dedicated search has been made looking for exhaust structures in these data sets, but absence of reports could suggest that exhaust structures sufficiently large to penetrate the ionosphere are, as expected, fairly rare. A search for expected signals in radar data sets could help answer this question.

\section{Conclusion}

The importance of the proposed mechanism lies in the fact that it requires neither that an entire flux tube be evacuated in order to perforate the ionosphere over a large altitude range, nor any sophisticated wave transformations or other complicated nonlinear effects. Evacuation is local and restricted to the extension of the electron exhaust structure along and perpendicular to the field. Transport of the radiation that is trapped in the exhaust is due to the movement of the structure along the geomagnetic field at the Alfv\'en speed of the IAW carrier wave. 
  
It is rather difficult to imagine any other mechanism than reconnection for the downward transport of X-mode AKR to altitudes as low as ground-level. Limited numbers of possibilities come to mind, such as are statistically rare large amplitude fluctuations in the ionospheric plasma density, large amplitude mirror modes, Kelvin-Helmholtz vortices and Rayleigh-Taylor density depletions. Each of them is not impossible but improbable. Large amplitude statistical depletions should occur rarely and even more so if required to occur during active AKR. Although AKR is almost continuously present somewhere around the globe, it would have to be in the right location to become conducted to low altitude by such depletions. Mirror modes require a perpendicular temperature anisotropy in the ion component, and low amplitude low-frequency whistler noise observed inside mirror modes \citep{baumjohann1999} show that the mirror-mode trapped electron distribution has indeed some weak temperature anisotropy. However, there is little ground to assume that the mirror mode trapped electrons are even only weakly relativistic, and non-relativistic temperature anisotropies must be extremely large for excitation of ECMI \citep{melrose1976}. The only possibility would be that the mirror modes trap some part of AKR when evolving while do not produce the ECMI themselves. Another problem is explaining why they fall down into the F and E layers and possibly even below the E-layer, when a low density bubble should rather rise upward. A similar argument applies to Rayleigh-Taylor bubbles which might evolve in the auroral jet current at F-layer altitudes. Trapping of AKR at these altitudes requires already downward transport of radiation which is not provided by the Rayleigh-Taylor modes themselves. Such bubbles do always rise only. Similarly, the Kelvin-Helmholtz instability can be driven by the shear flows which evolve between adjacent regions of upward and downward current regions at upper auroral region altitudes, a well established process causing Kelvin-Helmholtz vortices which are accompanied by density depletions and compressions and are seen as vortices in active aurorae. Such vortices may move both up and down, with no easily predictable  preferred direction. Since no reason is known for how the depletions could generate AKR via the ECMI, the only reasonable assumption would again be that the depletions trap radiation and transport it down when sinking. In summary, none of these alternative mechanisms have the key advantages of the mechanism proposed herein: generation of X-mode AKR naturally within the structures, jetting of the structures at the Alfv\'en velocity from the magnetosphere toward low altitudes, transporting radiation with them, and the possibility for the largest of the structures to survive the trip and stay sufficiently large to contain the trapped radiation, releasing it only at low altitudes possibly even below the ionosphere. 

The resulting low-altitude AKR should be impulsive, quasi-periodic, and might have complex frequency-time structure. Most observations have insufficient time resolution to check for these characteristics. The few that have sufficient time resolution show complex frequency time structure but neither impulsive nor quasi-periodic properties. However, these observations are so few in number, they do not disprove the mechanism described above, though they do suggest that if it occurs, the generation and trapping of X-modes in reconnection exhaust structures can only explain some, not all, low-altitude AKR events. Possibly, most events are explained by mechanisms which completely evade transport of X-mode to low altitude, such as mode conversion of ECMI-generated waves to whistler modes near their sources followed by ducted or non-ducted propagation to low altitudes. The mechanism proposed herein is nevertheless important to confirm, even if it explains only a minority of examples, because it would provide another effective means to remotely sense the extremely important process of magnetospheric magnetic reconnection located in the auroral region, and it would provide needed experimental evidence for the electron exhaust structures seen in PIC simulations of such reconnection to also occur under auroral conditions in the upper ionosphere which is required to explain the observed AKR fine structure as caused by the ECMI. In view of application to radiation from stars, planets, and astrophysical turbulence, confirmation as well as rejection of reconnection as a source of the ECMI is of vital importance. 

\begin{acknowledgement}
This work was part of a Visiting Scientist Programme at the International Space Science Institute Bern. We acknowledge the interest of the ISSI Directors Rafael Rodrigo and Rudolf von Steiger and the hospitality of the ISSI staff. We thank the ISSI system administrator Saliba F. Saliba for technical support and the librarians Andrea Fischer and Irmela Schweizer for access to the library and literature. LB acknowledges the substantial technical contributions of Mike Trimpi (Dartmouth College) to the success of the South Pole experiment. The work at Dartmouth College was supported by grant PLR-1443338 from the U.S. National Science Foundation.
\end{acknowledgement}

\noindent\emph{Data availability.} No data sets were used in this article.

\noindent\emph{Competing interests.} The authors declare that they have no conflict of interest.

\end{document}